\documentstyle[12pt]{article}
\begin{document}
\baselineskip6mm
\title{\vspace{-3cm} Unified gauge models and one-loop quantum
cosmology}
\author{G. Esposito$^{1,2}$, \ A. Yu. Kamenshchik$^{3}$ and
\ G. Miele$^{2,1}$}
\date{}
\maketitle
\hspace{-6mm}$^{1}${\em Istituto Nazionale di Fisica Nucleare,
Sezione di Napoli, Mostra d'Oltremare Padiglione 20, 
80125 Napoli, Italy}\\
$^{2}${\em Dipartimento di Scienze Fisiche, Mostra
d'Oltremare Padiglione 19, 80125 Napoli, Italy}\\
$^{3}${\em Nuclear Safety Institute, Russian
Academy of Sciences, 52 Bolshaya Tulskaya, Moscow 113191, Russia}\\

\begin{abstract} This paper studies the normalizability
criterion for the one-loop wave function
of the universe in a de Sitter background, when various unified 
gauge models are considered. It turns out that, in the absence
of interaction between inflaton field and other matter fields,
the supersymmetric version of such unified models is preferred.
By contrast, the interaction of inflaton and matter fields, 
jointly with the request of normalizability at one-loop order,
picks out non-supersymmetric versions of unified gauge models.
\end{abstract}

The investigations in modern cosmology have been devoted to
two main issues. On one hand, there were the attempts to
build a quantum theory of the universe with a corresponding
definition and interpretation of its wave function 
\cite{1,2}. On the
other hand, the drawbacks of the cosmological standard model
motivated the introduction of inflationary scenarios. These 
rely on the existence of one or more scalar fields, and a
natural framework for the consideration of such fields is
provided by the current unified models of fundamental
interactions \cite{3}. The unification program started with 
the proposal and the consequent experimental verification of the 
electroweak standard model ($SU(3)_C \otimes SU(2)_L \otimes U(1)_Y$),
and has been extended to other simple 
gauge groups, like $SU(5)$, $SO(10)$ 
and $E_6$. All of them in fact, even if with different capability,
unlike the electroweak standard model 
are able to allocate all matter fields in a few 
irreducible representations (IRR) of the gauge group,
and require a small number of free parameters. However, since these
enlarged gauge models predict new physics, a first source of constraints
upon them is certainly provided by the experimental bounds on processes like
proton decay, neutrino oscillations, etc.. \cite{4}. Further restrictions
can be obtained from their cosmological applications, as
discussed in \cite{5}. 

One can say, however, that the majority of investigations,
studying the mutual relations between particle physics and
cosmology, leave quantum cosmology itself a bit aside, using
it only as a tool to provide initial conditions for inflation.
Meanwhile, one can get some important restrictions on particle
physics models, using general principles of quantum theory
such as normalizability of the wave function [6--12] or quantum
consistency of the theory \cite{13}. 

Our paper, following Refs. [6--12], studies the possible 
restrictions on unified gauge models resulting from a
one-loop analysis of the wave function of the universe and
from the request of its normalizability. It is known that
the Hartle-Hawking wave function of the universe \cite{1}, as 
well as the tunnelling one \cite{2}, are not normalizable at
tree level \cite{14}. In Ref. \cite{6} it was shown that, by taking 
into account the one-loop correction to the wave function,
jointly with a perturbative analysis of cosmological 
perturbations at the classical level, one can obtain a
normalizable wave function of the universe provided that
a restriction on the particle content of the model is
fulfilled. 

Such a restriction is derived from the formula for the
probability distribution for values of the inflaton
field \cite{6}
\begin{equation}
\rho_{HH,T}(\varphi) \cong {1\over H^{2}(\varphi)}
{\mbox e}^{\mp I(\varphi)
-\Gamma_{1-\rm{loop}}(\varphi)} \; ,
\end{equation}
where HH and T denote the Hartle-Hawking and tunnelling wave
function, respectively, $H(\varphi)$ is the effective Hubble
parameter, $\Gamma_{1-\rm{loop}}$ is the one-loop effective
action on the compact de Sitter instanton. One can show from (1),
that the normalizability condition of the probability 
distribution at large values of the inflaton scalar field
$\varphi$ is reduced to the condition \cite{6}
\begin{equation}
Z > -1 \; ,
\end{equation}
where $Z$ is the total anomalous scaling of the theory. This
parameter is determined by the total Schwinger-DeWitt
coefficient $A_{2}$ in the heat-kernel asymptotics \cite{15},
and depends on the particle content.

In Ref. \cite{8} the criterion (2) was used to investigate the
permissible content of different models. It was noticed
that the standard model of particle physics, as well as
the minimal $SU(5)$ GUT model, does not satisfy the criterion
of normalizability, while the standard supersymmetric model,
the $SU(5)$ SUSY model and $SU(5)$ supergravity model do
satisfy this criterion. 

All the analysis in Ref. \cite{8} was carried out in terms of
physical degrees of freedom, e.g. 3-dimensional transverse
photons or 3-dimensional transverse-traceless metric
perturbations. However, over the last few years, the explicit
calculations have shown that a covariant path integral for
gauge fields and gravitation yields an anomalous scaling which
differs from the one obtained from reduction to physical
degrees of freedom. For compact manifolds without boundary
this discrepancy can be appreciated by comparing the results
of Ref. \cite{16} and Ref. \cite{17}. For manifolds with boundary we
refer the reader to the work in Refs. \cite{18,19} and references
therein.

Unfortunately, the reduction to physical degrees of freedom
relies on a global foliation by three-dimensional 
hypersurfaces which is only well-defined when the Euler
number of the four-dimensional Riemannian manifold vanishes.
Moreover, such a reduction does not take explicitly into
account gauge and ghost terms in the path integral, and leads
to a heat-kernel asymptotics which disagrees with the
well-known results of invariance theory 
\cite{17,20}. For all these
reasons, we regard the covariant version of the path integral
as more appropriate for one-loop calculations.

In Ref. \cite{9} the investigation of the one-loop wave function
was carried out for a non-minimally coupled inflaton field
with large negative constant $\xi$. It was then shown that 
the behaviour of the total anomalous scaling $Z$ is determined
by interactions between the inflaton and remaining matter
fields. 

Here, we study normalizability properties of a wide set of
unified gauge models, with or without interaction with the
inflaton field. The models studied are, as shown in Table I,
the standard model of particle physics, $SU(5)$, $SO(10)$
model in the 210-dimensional irreducible representation,
$E_{6}$, jointly with supersymmetric versions of all these
models with or without supergravity. The building blocks of
our one-loop analysis are the evaluations of $A_{2}$ coefficients
for scalar, spinor, gauge, graviton and gravitino perturbations.
All these coefficients (but one) are, by now, well-known
(e.g. \cite{20,21}) and are given by
\begin{eqnarray}
A_{2 \; \rm {scalar}}&=&
{29\over 90}-4\xi+12 \xi^{2} \nonumber\\
&-& {1\over 3}m^{2}R_{0}^{2}
+2\xi m^{2} R_{0}^{2}
+{1\over 12}m^{4}R_{0}^{4} \; ,
\end{eqnarray}
\begin{equation}
A_{2 \; {\rm {spin}}-1/2}={11\over 180}+{1\over 3}
m^{2}R_{0}^{2}+{1\over 6}m^{4}R_{0}^{4} \; ,
\end{equation}
\begin{equation}
A_{2 \; \rm {gauge}}=-{31\over 45}+{2\over 3}m^{2}
R_{0}^{4}+{1\over 3}m^{4}R_{0}^{4} \; ,
\end{equation}
\begin{equation}
A_{2 \; \rm {gravitino}}=-{589\over 180} \; .
\end{equation}
It should be stressed that Eq. (3) only holds for scalar
fields different from the inflaton. With our notation,
$m,\xi$ and $R_{0}$ represent effective mass, (dimensionless)
coupling parameter, and 4-sphere radius, respectively.
Equation (4) describes a spin-1/2 field with half the number
of modes of a Dirac field. Since the results (5) and (6) rely
on the Schwinger-DeWitt technique, they incorporate, by
construction, the effect of ghost zero modes. However, it has
been argued in Ref. \cite{22} that zero modes should be excluded 
to obtain an infrared finite effective action which is smooth
as a function of the de Sitter radius on spherically symmetric
backgrounds. On the other hand, the prescription which includes
ghost zero modes makes the one-loop results continuous. Strictly,
we are considering small perturbations of a de Sitter background
already at a classical level (see [6--12]). There are also deep
mathematical reasons for including zero modes, and they result
from the spectral theory of elliptic operators \cite{23}. Thus, we
use the expressions (5) and (6).

Last, the contribution of gravitons to the total $Z$ should be
calculated jointly with the inflaton contribution. What happens
is that the second-order differential operator given by the
second variation of the action with respect to inflaton and metric
is non-diagonal even on-shell, by virtue of a non-vanishing
vacuum average value of the inflaton \cite{24,25}. The resulting
$A_{2}$ coefficient turns out to be independent of the value
of $\xi$ and equal to \cite{12}
\begin{equation}
A_{2 \; {\rm graviton} + {\rm inflaton}}=-{171\over 10} \; .
\end{equation}
In the following table, we report the total $Z$ for some relevant
examples of GUT theories, whenever one neglects the 
mass terms. This ansatz is correct, if the interaction
between inflaton and the other particles is not considered. In this
case in fact, the term $m^2 R_0^2 \sim\varphi^{-2}$ is very small
due to the large value of $\varphi$. The analysis starts with the 
electroweak standard model (SM), which contains, in its non-SUSY
version, $45$ Weyl spinors (we neglect for simplicity right-handed
neutrinos and their antiparticles), $24$ gauge bosons and one doublet 
of complex Higgs fields. The particle content changes for the SUSY
version of this model in its minimal form (MSSM) \cite{26}. 
In this case, in fact, 
to the $45$ Weyl leptons and quarks one has to add $4$ higgsinos 
and $12$ gauginos, whereas the scalar sector consists now of 
$90$ sleptons and squarks plus $8$ real scalar fields. 
A similar analysis is performed for the
$SU(5)$ GUT model \cite{27},
which in its non-SUSY version, apart from the $24$
gauge bosons, needs scalars belonging to $\underline{24} \oplus 
\underline{5} \oplus \underline{\overline{5}}$ IRR's to accomplish 
the spontaneous symmetry breaking pattern. 
The matter content of the SUSY extension of the 
model \cite{28} is obtained by 
doubling the number of Higgs IRR's used, and by adding superpartners
to any degrees of freedom. 
As far as $SO(10)$ gauge theories are concerned, 
we have considered the particular model containing 
$\underline{210} \oplus (\underline{126} \oplus \underline{\overline{126}})
\oplus \underline{10} \oplus \underline{10}$ IRR's of Higgs fields,
which is still compatible with the present experimental limit on the 
proton lifetime and neutrino phenomenology \cite{4}. Furthermore, we have 
also considered the SUSY extension of $SO(10)$, which, to be 
consistent also with cosmological constraints, needs complex   
Higgs fields belonging to $\underline{1} \oplus \underline{10}
\oplus \underline{10}' \oplus \underline{45} 
\oplus \underline{45}' \oplus \underline{54} \oplus
\underline{54}' \oplus \underline{126} \oplus \underline{126}'$
IRR's \cite{29}.

\footnotesize
\bigskip\bigskip
\par\noindent
{\bf Table I.}\\
\bigskip
\begin{tabular}{|c|c|c|c|}
\hline
 & & & \\
Gauge group & version & $Z$ & forbidden $\xi$ range \\
 & & & \\  
\hline   
& & & \\
& non-SUSY & $36 \xi^2 -12 \xi - {543 \over 20}$ & 
$-.701 \leq \xi \leq 1.035$ \\
& & & \\  
$SU(3)_C \otimes SU(2)_L \otimes U(1)_Y$ & SUSY & 
$1164 \xi^2 -388 \xi + {389 \over 180}$ &  
$.008 \leq \xi \leq .325$ \\
& & & \\  
& SUGRA & $1164 \xi^2 -388 \xi +{163 \over 30}$ & 
$.017 \leq \xi \leq .316$ \\   
& & & \\  
\hline 
& & & \\  
& non-SUSY & $ 396 \xi^2 - 132 \xi -{103 \over 4} $ & 
$-.134 \leq \xi \leq .467 $ \\
& & & \\  
$SU(5)$ & SUSY & 
$1884 \xi^2 - 628 \xi +{1919 \over 180} $ &  
$.020 \leq \xi \leq .314$ \\
& & & \\  
& SUGRA & $ 1884 \xi^2 -628 \xi +{209 \over 15}$ & 
$ .026 \leq \xi \leq .308$ \\   
& & & \\
\hline 
& & & \\
& non-SUSY & $ 5772 \xi^2 - 1924 \xi + {4678 \over 45} $ & 
$.069 \leq \xi \leq .265 $ \\
& & & \\  
$SO(10)$ & SUSY & 
$ 12444 \xi^2 - 4148 \xi + {11321 \over 45} $ &  
$.080 \leq \xi \leq .253$ \\
& & & \\  
& SUGRA & $ 12444 \xi^2 - 4148 \xi +{5097 \over 20}$ & 
$.082 \leq \xi \leq .252$ \\   
& & & \\  
\hline 
& & & \\  
& non-SUSY & $ 10932 \xi^2 -3644 \xi +{39197 \over 180} $ & 
$.078 \leq \xi \leq .255 $ \\
& & & \\  
$E_6$ & SUSY & 
$ 12876 \xi^2 - 4292 \xi +{42719 \over 180} $ &  
$.070 \leq \xi \leq .263$ \\
& & & \\  
& SUGRA & $ 12876 \xi^2 -4292 \xi +{1203 \over 5}$ & 
$.072 \leq \xi \leq .262$ \\   
& & & \\
\hline 
\end{tabular}
\normalsize   
\baselineskip6mm   
Finally, we have also considered $E_6$ GUT theories, for which
fermions are allocated in three $\underline{27}$ fundamental IRR's,
and scalars belong to two $(\underline{78} \oplus \underline{27}   
\oplus \underline{351})$ \cite{30}. For the SUSY extension of
this model, we have just added the superpartner degrees of freedom.
Concerning the SUGRA versions of all the above models, they
have been obtained from the supersymmetric ones, just by adding 
the gravitino contribution (i.e. subtracting the $A_{2}$
coefficient in Eq. (6), because of the fermionic statistics).
Indeed, we have considered particular versions of $SO(10)$ and
$E_{6}$ gauge models, but we expect that the qualitative 
features of the results (see below) should remain unaffected.

In Table I, we have assumed that one of the Higgs fields plays
the role of the inflaton. 
The forbidden range denotes the range of values
of $\xi$ for which the normalizability criterion (2) is not
satisfied. Interestingly, conformal coupling 
(i.e. $\xi=1/6$) is ruled out
by all 12 models listed in Table I. Moreover, for the
standard and $SU(5)$ models, minimal coupling (i.e.
$\xi=0$) is also ruled out. At this stage, supersymmetric
models are hence favoured, as well as non-supersymmetric
models with a large number of scalar fields.

However, realistic cosmological models should include 
interactions between the inflaton and remaining fields.
Hence these fields acquire masses proportional to the 
inflaton vacuum average value, i.e. proportional to
$R_{0}^{-1}$. In such a way, mass terms in Eqs. (3)--(5)
become independent of $R_{0}$. Moreover, the consideration
of cosmological perturbations shows that $m^{2}R_{0}^{2}$ is
of order $10^{4}$ \cite{31}. Thus, these terms provide the
dominant contribution to the anomalous scaling factor.
In non-supersymmetric unified models, one can switch on
interactions between scalar fields and the inflaton without
any further interaction with spinor fields. In this case,
the occurrence of positive terms proportional to 
$m^{4}R_{0}^{4}$ in Eq. (3) ensures the fulfillment of the one-loop
normalizability criterion (2). However, if one considers
supersymmetry, jointly with a Wess-Zumino scalar multiplet
interacting with the inflaton, the terms proportional to
$m^{4}R_{0}^{4}$ in Eqs. (3) and (4) cancel each other exactly.
By contrast, terms proportional to $m^{2}R_{0}^{2}$ are large
but have opposite signs, so that their combined effect makes
it impossible to satisfy the condition (2). Thus, the naive
argument in favour of supersymmetry presented in Ref. \cite{8}
and supported by our Table I, fails whenever inflaton 
interactions are taken into account. 

Despite the arguments which seem to rule out a
class of supersymmetric models as described so far, our
investigation cannot really be used to discriminate
supersymmetry at this stage. We have just combined 
the various contributions to the $A_{2}$
coefficient resulting from the particle content of the
gauge models under consideration. No systematic investigation
of supersymmetric quantum cosmology, however, has been
presented, following for example the Hamiltonian analysis of
Ref. \cite{32} (see also Ref. \cite{33}). In other words, the thorough
consideration of supersymmetry constraints and auxiliary fields
along the lines suggested in Ref. \cite{32} might provide another
(and possibly deeper) approach to the inclusion of supersymmetric
gauge models in the analysis of (one-loop) quantum cosmology.

A. K. is grateful to A.O. Barvinsky for useful discussions.
A. K. is indebted to the Istituto Nazionale di Fisica
Nucleare for financial support, and to the Dipartimento di
Scienze Fisiche of the University of Naples for hospitality.
A. K. was partially supported by RFBR via grant no. 
96-02-16220 and RFBR-INTAS via grant no. 644, and by Russian
research project ``Cosmomicrophysics".


\begin{thebibliography}{99}
\bibitem{1}
J.B. Hartle and S.W. Hawking, Phys. Rev. D 28 (1983) 2960.
\bibitem{2}
A. Vilenkin, Phys. Rev. D 37 (1988) 888.
\bibitem{3}
A.D. Linde, Particle physics and inflationary cosmology
(Harwood Academic, New York, 1990).
\bibitem{4}
F. Acampora, G. Amelino-Camelia, F. Buccella, O. Pisanti,
L. Rosa and T. Tuzi, Nuovo Cimento A 108 (1995) 375.
\bibitem{5}
G. Esposito, G. Miele and P. Santorelli, Phys. Rev.
D 54 (1996) 1359.
\bibitem{6}
A.O. Barvinsky and A.Yu. Kamenshchik, Class. Quantum
Grav. 7 (1990) L181.
\bibitem{7}
A.O. Barvinsky, Phys. Rep. 230 (1993) 237.
\bibitem{8}
A.Yu. Kamenshchik, Phys. Lett. B 316 (1993) 45.
\bibitem{9}
A.O. Barvinsky and A.Yu. Kamenshchik, Phys. Lett. B
332 (1994) 270.
\bibitem{10}
A.O. Barvinsky and A.Yu. Kamenshchik, Phys. Rev. D 50
(1994) 5093.
\bibitem{11}
A.O. Barvinsky, Phys. Rev. D 50 (1994) 5115.
\bibitem{12}
A.O. Barvinsky, A.Yu. Kamenshchik and I.V. Mishakov,
Quantum origin of the early inflationary universe, 
submitted to Nucl. Phys. B.
\bibitem{13}
A.Yu. Kamenshchik and S.L. Lyakhovich, Hamiltonian
BFV-BRST theory of closed quantum cosmological models,
hep-th/9608130.
\bibitem{14}
S.W. Hawking and D.N. Page, Nucl. Phys. B 264 (1986) 185.
\bibitem{15}
A.O. Barvinsky and G.A. Vilkovisky, Phys. Rep. 119 (1985) 1.
\bibitem{16}
P.A. Griffin and D.A. Kosower, Phys. Lett. B 233 (1989) 295.
\bibitem{17}
D.V. Vassilevich, Phys. Rev. D 52 (1995) 999.
\bibitem{18}
G. Esposito, A.Yu. Kamenshchik, I.V. Mishakov and 
G. Pollifrone, Class. Quantum Grav. 11 (1994) 2939.
\bibitem{19}
G. Esposito, A.Yu. Kamenshchik, I.V. Mishakov and
G. Pollifrone, Phys. Rev. D 52 (1995) 3457.
\bibitem{20}
P.B. Gilkey, Invariance theory, the heat equation, and
the Atiyah-Singer index theorem (Chemical Rubber Company,
Boca Raton, 1995).
\bibitem{21}
I.G. Moss, Quantum theory, black holes and inflation
(Wiley, New York, 1996).
\bibitem{22}
T.R. Taylor and G. Veneziano, Nucl. Phys. B 345 (1990) 210.
\bibitem{23}
M.F. Atiyah, V.K. Patodi and I.M. Singer, Math. Proc.
Camb. Phil. Soc. 79 (1976) 71.
\bibitem{24}
Yu.V. Gryzov, A.Yu. Kamenshchik and I.P. Karmazin,
Izv. VUZov, Fiz. (Russia) 35 (1992) 121.
\bibitem{25}
A.O. Barvinsky, A.Yu. Kamenshchik and I.P. Karmazin,
Phys. Rev. D 48 (1993) 3677.
\bibitem{26} 
P. Fayet, Phys. Lett. B 64 (1976) 159.
\bibitem{27}
H. Georgi and S.L. Glashow, Phys. Rev. Lett. 32 
(1974) 438.
\bibitem{28}
S. Dimopoulos and H. Georgi, Nucl. Phys. B 193 (1981) 150.
\bibitem{29} 
R. Jeannerot, Phys. Rev. D 53 (1996) 5426.
\bibitem{30} 
J.L. Hewett and T.G. Rizzo, Phys. Rep. 183 (1989) 193.
\bibitem{31}
D.S. Salopek, J.R. Bond and J.M. Bardeen, Phys. Rev.
D 40 (1989) 1753.
\bibitem{32}
P.D. D'Eath, Supersymmetric quantum cosmology (Cambridge
University Press, Cambridge, 1996).
\bibitem{33}
P.R.L.V. Moniz, Supersymmetric quantum cosmology: shaken
not stirred, gr-qc/9604025.
\end{thebibliography}
\end{document}